# Health Wearables, Gamification, and Healthful Activity

Muhammad Zia Hydari,[a,]* Idris Adjerid,[b,]* Aaron D. Striegel[c]

[a] Katz Graduate School of Business, University of Pittsburgh, Pittsburgh, Pennsylvania 15260; [b] Pamplin College of Business, Blacksburg, Virginia 24061; [c] Department of Computer Science and Engineering, University of Notre Dame, Notre Dame, Indiana 46556
*Corresponding author
Contact: hydari@alum.mit.edu, https://orcid.org/0000-0003-4522-326X (MZH); iadjerid@vt.edu, https://orcid.org/0000-0002-2786-1244 (IA); striegel@nd.edu, https://orcid.org/0000-0002-3157-2859 (ADS)

**Abstract.** Health wearables in combination with gamification enable interventions that have the potential to increase physical activity—a key determinant of health. However, the extant literature does not provide conclusive evidence on the benefits of gamification, and there are persistent concerns that competition-based gamification approaches will only benefit those who are highly active at the expense of those who are sedentary. We investigate the effect of Fitbit leaderboards on the number of steps taken by the user. Using a unique data set of Fitbit wearable users, some of whom participate in a leaderboard, we find that leaderboards lead to a 370 (3.5%) step increase in the users' daily physical activity. However, we find that the benefits of leaderboards are highly heterogeneous. Surprisingly, we find that those who were highly active prior to adoption are hurt by leaderboards and walk 630 fewer steps daily after adoption (a 5% relative decrease). In contrast, those who were sedentary prior to adoption benefited substantially from leaderboards and walked an additional 1,300 steps daily after adoption (a 15% relative increase). We find that these effects emerge because sedentary individuals benefit even when leaderboards are small and when they do not rank first on them. In contrast, highly active individuals are harmed by smaller leaderboards and only see benefit when they rank highly on large leaderboards. We posit that this unexpected divergence in effects could be due to the underappreciated potential of noncompetition dynamics (e.g., changes in expectations for exercise) to benefit sedentary users, but harm more active ones.

**History:** Accepted by Chris Forman, information systems.

**Open Access Statement:** This work is licensed under a Creative Commons Attribution 4.0 International License. You are free to copy, distribute, transmit and adapt this work, but you must attribute this work as "*Management Science*. Copyright © 2022 The Author(s). https://doi.org/10.1287/mnsc.2022.4581, used under a Creative Commons Attribution License: https://creativecommons.org/licenses/by/4.0/."
**Funding:** This work was supported by the National Institutes of Health [Grant 5R01HL117757].
**Supplemental Material:** The data and online appendix are available at https://doi.org/10.1287/mnsc.2022.4581.

## 1. Introduction

The evidence on the health[1] benefits of physical activity is irrefutable (Warburton et al. 2006). Yet, a significant portion of the world population is not sufficiently active.[2] This lack of physical activity contributes significantly to chronic disease and to most of the leading causes of death in the United States.[3] Prior research suggests that behavioral barriers are one of the most important contributing factors to this trend. Mitchell et al. (2013) suggest that "for many adults, the 'costs' of exercise (e.g., time, uncomfortable feelings) loom so large that they never start" and that the lack of physical activity is "a problem of both initiation and maintenance" (p. 658). Recognizing that changing health behaviors is often challenging and new strategies are needed, research situated mostly in the health and economics literature has evaluated a plethora of economic and non-economic interventions for overcoming motivational barriers to increasing physical activity (Charness and Gneezy 2009, Mitchell et al. 2013). The conclusion from this literature is that, although many interventions can drive short-run gains in physical activity, these benefits are fleeting and motivating meaningful, and sustained increases in physical activity is elusive. More so, many of these interventions (e.g., daily payments) are difficult to implement on a population scale.

One contemporary phenomenon with the potential to address persistent limitations of prior approaches and unlock new interventions that can improve the motivation of individuals to exercise is the rapid consumer adoption of health wearables (Swan 2013, Lupton 2016). A health wearable, sometimes referred to as an activity tracker, is "a wearable device or a computer application that records a person's daily physical activity, together with other data relating to their fitness or health, such as the number of calories burned, heart rate, etc."[4] Despite the rapid adoption of health wearables and their potential for motivating individuals to engage in healthy activities, scholars suggest it is unlikely that the measurement capabilities that the health wearables provide would significantly impact health on their own (Patel et al. 2015, Sullivan and Lachman 2017). Rather, they suggest that for health wearables to impact health behavior, the information they collect "must be presented back to the user in a manner that can be understood, that motivates action, and that sustains that motivation toward improved health" (Patel et al. 2015, p. 460).

Particularly promising in this regard is combining granular physical activity data from health wearables with gamification approaches. Gamification is defined as the "use of game design elements in nongame contexts" (Deterding et al. 2011). Some examples of game design elements are badges, rules-based competition, leaderboards, points, ranking, reputation, rewards, teams, and time pressure (Deterding et al. 2011). Coupling gamification with health wearables has the potential to improve motivation by converting the usually mundane action of physical activity into the more enjoyable activity of collecting rewards or competing with other individuals (Hamari et al. 2014a, Johnson et al. 2016). More so, gamification approaches can provide immediate positive reinforcement that helps individuals get over the initial hurdles of engaging in exercise and could also help them sustain higher levels of activity in the longer term (Mitchell et al. 2013, Shameli et al. 2017). In addition, the broad adoption of health wearables unlocks more robust gamification interventions by providing an objective and common source of measurement and a form factor that enables real-time feedback while engaging in physical activity (Johnson et al. 2016).

Although coupling gamification with health wearables has the potential to generate sustained increases in physical activity, the evidence on the benefits of gamification is mixed. Hamari et al. (2014b) reviewed 24 empirical gamification studies, primarily within education contexts, but reported mixed effects on outcomes. Moreover, these studies used interviews or surveys to measure outcomes. Hamari and Koivisto (2013), the only empirical study in a health context in the aforementioned literature review, used surveys to measure outcomes and these outcomes were *not* health related (e.g., continued use intention for the gamification service and the intention to recommend service to others). Johnson et al. (2016) conducted a systematic literature review on the impact of gamification on health and well-being. Of the 19 empirical papers they reviewed, 59% reported positive results and 41% reported mixed results. Both Hamari et al. (2014b) and Johnson et al. (2016) also noted that the quality of evidence was moderate to low.

The significant potential benefits of coupling gamification with health wearables and the narrow focus and lack of evidentiary quality of prior works motivate this research study. We evaluate the benefits of leaderboards that allow users to view the performance of others who also agree to share their activity levels and, in most cases, to compete with them. We focus on the potential benefits of leaderboards because they are one of the most common gamification features available with modern health wearables. This increases the policy and practical relevance of our results. Another reason we focus on leaderboards is that they exemplify the theoretical tensions surrounding gamification interventions; scholars suggest that gamification features, and leaderboards in particular, are likely to have heterogeneous effects on individuals (Deci et al. 1981, Santhanam et al. 2016, Sullivan and Lachman 2017). Specifically, a central concern with competition-based gamification interventions like leaderboards is that they will lead to motivational benefits only for those who are already highly active (and need the increased motivation the least), whereas actually harming the least physically active in the population (Patel et al. 2015, Wu et al. 2015, Shameli et al. 2017).[5] With these dynamics in mind, our first objective is to evaluate the average impact on physical activity of leaderboard adoption by individuals wearing health wearables. Our second research objective is to evaluate the potential for leaderboard effect heterogeneity by (i) the activity level of the focal user prior to adoption, (ii) the number of active participants on the leaderboard, and (iii) the rank of the focal user on the leaderboard in the prior period.

We engage in an intensive data collection effort to estimate the average benefit of leaderboards and the heterogeneity in these benefits. For approximately 500 individuals observed over a two-year time period, we capture leaderboard adoption data and granular measures of physical activity continuously captured by Fitbit Charge HR health wearables. For those individuals with leaderboards, the data set also includes activity data and rank of all participants in the leaderboard. We supplement these data from health wearables with periodic surveys (every six months on average) capturing a rich array of individual characteristics (psychological attributes, frequency of technology use, etc.). Leveraging variation in physical activity and adoption of leaderboards over time and between individuals, we use a difference-in-differences (DID) estimation approach to evaluate the effect of leaderboards on daily physical activity as measured by the user's step count, as well as heterogeneity in these effects.

We find that leaderboard adoption results in an average daily increase of 370 steps. This main effect is resilient to various tests for the assumption of common trends between those who adopt and do not adopt, estimation of several falsification tests, and other robustness checks. These initial results, however, mask important heterogeneity in the benefits of leaderboard adoption. When we take into account an individual's prior activity levels, we find a stark divergence in leaderboard effects. Individuals who were highly active prior to adoption, instead of benefiting from leaderboards, experienced a significant decrease in their average daily physical activity after leaderboard adoption (a decrease of 631 steps daily). Moreover, these negative effects persisted (and actually increased in magnitude) in the 10 weeks following leaderboard adoption. In contrast, users who were less active prior to adoption had large and significant positive impacts on their daily step counts—On average, their activity increased by 1,365 steps daily (an approximately 15% increase), and these increases also persisted well after the adoption decision (10 weeks after adoption).

Examining this trend further, we find significant nuance in how leaderboard size impacts sedentary versus highly active users. Specifically, the key distinction between these groups is that previously sedentary individuals can reap significant benefit from small leaderboards (only one or two other members) and even if they do not rank first. In contrast, those who were highly active (prior to adoption) see the most significant harm when leaderboards are small. Our interpretation of these results is that individuals who are already on the high end of the physical activity distribution can become complacent on small leaderboards where, more often than not, they are paired with those less active than themselves.[6] In contrast, sedentary individuals who are at the lower end of the distribution of physical activity often encounter (even on small leaderboards) peers who are more active than themselves, who can positively impact their reference point for exercise, and who can hold them accountable if their activity levels slump. However, these benefits for sedentary users diminish if leaderboards become too large; the marginal benefit of an additional leaderboard member diminishes at three times the rate for sedentary users relative to highly active ones. One explanation for this effect is that the benefits of social influence that accrue to sedentary users (e.g., positive impact on their exercise reference points) diminish as leaderboards become larger and less intimate.

Our research contributes to streams of work at the intersection of information systems, economics, and healthcare. Specifically, we contribute to the literature on the economics of health information technology (IT) and specifically to the nascent streams of work evaluating economic and health implications of widespread adoption of health wearables (Handel and Kolstad 2017) and the broader potential of digital platforms to unlock interventions that leverage social norms and reciprocity to improve health (Liu et al. 2019a; b; Sun et al. 2019). Currently, the evidence on benefits from health wearables does not align with their promise. Piwek et al. (2016, p. 2) suggest that "current empirical evidence is not supportive" of health benefits from health wearables. Recent studies using large samples and robust causal approaches find little or no benefit on health outcomes of using health wearables (Lewis et al. 2015, Finkelstein et al. 2016, Jakicic et al. 2016). However, scholars have argued that a limitation of prior works is that they do not adequately consider the role of innovative technology decision aids and behavioral interventions enabled by broad adoption of health wearables (Patel et al. 2015). Our study addresses these limitations of prior work and finds that, on average, leaderboards promote healthful activity. However, our results also caution that these benefits may be highly nuanced with considerable variation in gains. In some cases, individuals may opt into variants of these interventions with either no benefit to them or, in some cases, negative effects on their physical activity.

We also contribute to the behavioral economics and information systems (IS) literature on gamification, especially, within the healthcare context. Despite mixed evidence of benefits and numerous open empirical and theoretical questions (Liu et al. 2017, Treiblmaier et al. 2018, James et al. 2019), gamification is spreading into a number of decision contexts. For instance, two recent IS papers have examined the impact of gamification within the retail context (Pamuru et al. 2021, Ho et al. 2022). Our study is differentiated with extant literature in several ways. First, our study is an individual-level intervention within healthcare in which the combination of unique data and rigorous estimation approaches results in more conservative estimates of average treatment effects of leaderboards; prior work showing positive effects of similar gamification interventions has found treatment effects five times our estimates (Shameli et al. 2017). Second, our results suggest that the mixed evidence of prior work may be explained, in part, by significant heterogeneity in gamification impacts. Not only are we able to provide more nuance in our study for gamification's impact (Ho et al. 2022), we also provide evidence on a substantively important issue in the medical literature, that is, the impact on the previously less active users (Patel et al. 2015). In our setting, the relatively conservative estimates of the *average* effect of leaderboards mask robust heterogeneous effects that are large in magnitude, statistically significant, and persistent over time. These heterogeneous effects support our theoretical conjecture that competition and social influence are key mechanisms underlying leaderboard effects but also highlight that these mechanisms can result in unexpected motivational and de-motivational effects. Specifically, we identify a divergence of benefit for sedentary versus highly active users that is opposite to the expectation for competition-based gamification in the literature. These findings point to the underappreciated role of social influence benefiting sedentary users but harming more active ones. These results not only have significant managerial implications for firms in the health wearable and gamification spaces, but also for policy makers, healthcare entities, and employers interested in improving health.

## 2. Background

Physical activity is a key element of healthful living and is known to have significant health benefits (Penedo and Dahn 2005, Warburton et al. 2006). Our main outcome variable is Fitbit step counts and includes a variety of these healthful physical activities, such as jogging, running, walking, playing sports, climbing stairs, and so on. Moreover, daily step counts are key to Fitbit leaderboards, as rankings on leaderboards are determined exclusively by differences in the step counts of the participants of the leaderboard.

### 2.1. Leaderboards

A leaderboard is "a large board for displaying the ranking of the leaders in a competitive event."[7] In a digital setting, the leaderboard may be displayed on a mobile application or an online dashboard. In this study, we utilize health wearables made by Fitbit Inc., which is a pioneering firm in this market.[8] Using Fitbit's online dashboard or the mobile application, a Fitbit user can invite another user (or receive an invitation) to join a leaderboard. If there is mutual agreement between the users to participate, both users will appear on each other's leaderboards. Each leaderboard ranks participating users based on seven-day running tallies of their steps.[9] The step counts shown on the leaderboard are directly captured by the Fitbit device and are not manually entered by the users, thus avoiding the measurement errors that may result from self-reported activity data.

Figure 1 shows four leaderboards, with the focal user labeled at the lower left corner. Each leaderboard can have the same or different user composition. For instance, Ash and Todd are connected to Mary and to each other. Dave is only connected to Mary, and Mary is connected to all other users. The leaderboards also show the seven-day step count of each participating user. Users are assigned ranks on participating leaderboards based on their seven-day step count relative to other users on that leaderboard. For instance, Mary is ranked second on her own leaderboard, but she is ranked first on Ash's and Todd's leaderboards. Thus, Ash and Todd may be motivated to do better by seeing their lower rank on the leaderboard relative to Mary. Users get feedback according to their rank on their own leaderboard. Although Mary dominates the highest number of leaderboards, the feedback she gets is that she is ranked second on her own leaderboard and must strive harder to achieve a first rank. Leaderboard adoption is "sticky" on the Fitbit platform. To de-adopt, users have to go through cumbersome steps and hide themselves via privacy settings.

## 3. Effect of Leaderboards on Healthful Physical Activity

Whether leaderboards will increase or decrease healthful physical activity is not entirely clear as the effect is unlikely to be similar for all individuals. Leaderboards can produce an effect on an individual's physical activity primarily by altering this individual's *willingness* to engage in physical activity. Specifically, we conjecture that changes in willingness to engage in physical activity occur primarily due to the introduction of competitive dynamics, increased individual accountability, and altering an individual's reference point for their own activity levels.

### 3.1. Competition

Social comparison theory suggests that a fundamental mechanism through which individuals assess their own ability is through comparison with others (Festinger 1954). Competitiveness is one manifestation of the social comparison process and drives individuals to increase

**Figure 1.** (Color online) Fitbit Leaderboard Composition for Four Individuals

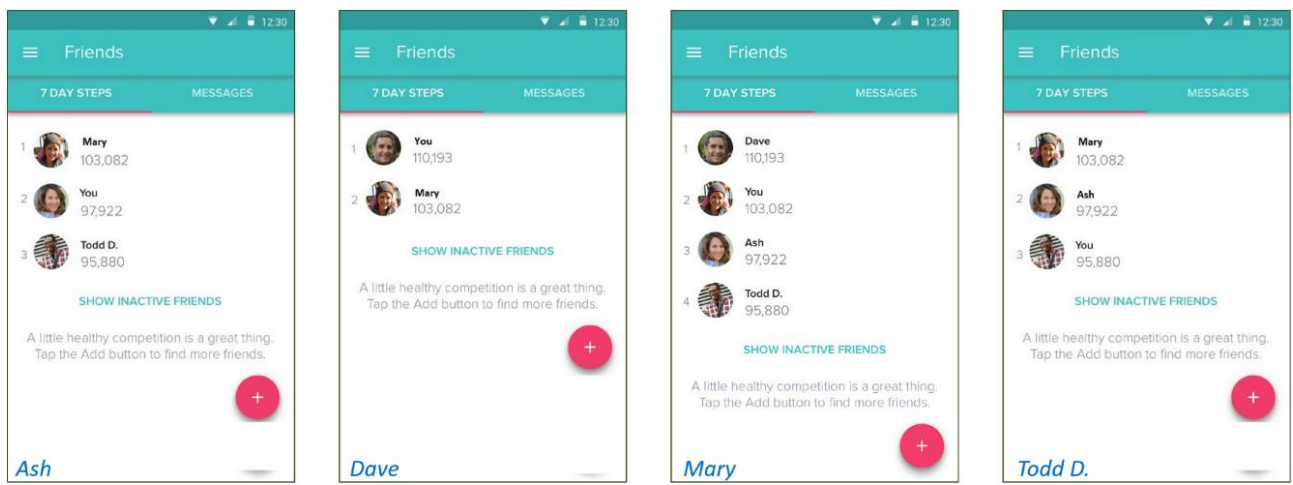

*Notes.* This figure shows leaderboards for Ash, Dave, Mary, and Todd. Ash and Todd are connected to each other and Mary, Dave is only connected to Mary, and Mary is connected to all other participants.

their effort either ex ante to elevate their rank or ex post to maintain their high rank (Garcia et al. 2013). Thus, the first and most direct way that leaderboards impact physical activity is through the competitive dynamic that ranking a focal user against other users generates. The tag line on the Fitbit leaderboard (Figure 1)—*a little healthy competition is a great thing*—points to the motivational potential of this competitive mechanism. In addition, the enjoyment derived from physical activity may be impacted by the individual's leaderboard adoption by converting the mundane activity of walking into the more exciting activity of competing against others. Therefore, individuals who may not gain any direct enjoyment from walking may engage in this activity because of the indirect enjoyment gained from competing on the leaderboard.

However, prior work finds that impacts of competition on motivation and effort are highly heterogeneous and depend on several factors such as a participant's desire to win, whether the competition provides a participant the opportunity or reason for improving their performance, and whether competition motivates a participant to put forth greater effort (Deci et al. 1981). Along this vein, a leaderboard may have minimal impact on performance if it does not provide sufficient competition or if the adopting individual is not particularly motivated by competition. More so, prior work has noted the possibility of competition having negative impacts on motivation and performance. For instance, Steinhage et al. (2015) argue that when competition elicits excitement, it may foster positive behavior. However, if competition elicits anxiety, it may foster negative behavior. Extrapolating this to our context, if the performance of others on the leaderboard elicits anxiety in the focal user, it would lead to negative outcomes for them. Reflecting this theoretical tension, the extant literature has found mixed results regarding competition with others who significantly surpass the individual in performance. Rogers and Feller (2016, p. 1) showed that "exposure to exemplary peer performances can undermine motivation and success by causing people to perceive that they cannot attain their peers' high levels of performance," and termed this phenomenon *discouragement by peer excellence*. However, Uetake and Yang (2019) find that an individual's distance from the highest achiever has positive motivational effects, whereas comparison with the average individual has negative impacts. Thus, competition is likely a focal mechanism behind leaderboard effects but whether it positively impacts physical activity is uncertain ex ante.

### 3.2. Social Influence

Leaderboards involve connecting individuals around health and the revelation of previously private levels of physical activity between individuals. These connections and disclosures introduce the potential of social influence to impact motivation and behavior. We consider two potential effects in the realm of social influence: individual accountability and reference points for exercise.[10]

**3.2.1. Individual Accountability.** Joining a leaderboard involves the revelation of one's previously private levels of physical activity to other users. This self-revelation allows other leaderboard members to hold the focal user accountable for lackluster levels of physical activity and nudge them to do better. In fact, the Fitbit app has a mechanism for messaging, cheering, and taunting other users directly from the platform. Some of these interactions may also happen off the Fitbit platform (and are thus unobserved by us as researchers)—for example, discussions between family members over dinner. The potential of group-based interventions to increase mutual accountability and increase physical activity has been explored in the literature: Patel et al. (2016) uncover benefits of incentive schemes for exercise that are tied to group versus individual performance targets.

**3.2.2. Exercise Reference Points.** In addition to the potential impacts of self-revelation, the revelation by others of their previously private levels of physical activity can result in changes to individuals' reference points for exercise. Specifically, social comparison theory suggests that such revelations can lead to an updated perception of one's own ability to exercise and the appropriateness of one's own level of exercise (Garcia et al. 2013). However, how these comparisons impact reference points depends on whether individuals engage in upward comparisons (i.e., comparisons with those more active than themselves) or downward comparisons (i.e., comparisons with those less active than themselves) (Festinger 1954). In both cases, the literature suggests that individuals will take action to reduce discrepancies between themselves and similar others (Festinger 1954, Garcia et al. 2013). Thus, if individuals compare upward, the revelation of this information between members of a leaderboard may have a positive impact on an individual's reference point for healthful activity and increase exercise. For instance, a mother with two young children may aim for a higher level of healthful activity if she observes another mother with two young children consistently doing more healthful activity. Given that these two users may have similar schedule constraints, the focal user may find the leaderboard information more relatable. If individuals compare downward, the revelation of physical activity information by others may have unintended negative impacts on an individual's reference point for physical activity. In particular, this informational signal can work in the opposite direction—that is, focal users may decrease activity if they see other relatable individuals on their leaderboards who are less active than themselves. Related to this point, Schultz et al. (2007)

found that a nudge intended to decrease electricity consumption by revealing the consumption levels of others in one's neighborhood had the (opposite) boomerang effect for those who were underusing electricity (relative to their neighbors) prior to the intervention.

### 3.3. Moderating Effects of Prior Activity Levels and Leaderboard Size

The contradictory effects of competition and social influence not only make the direction of the average effect uncertain, they also point to the presence of heterogeneity in the effects. To untangle this heterogeneity, we consider factors that can impact the propensity for observing the positive vs. negative dynamics of leaderboards on physical activity.

**3.3.1. Leaderboard Size.** First, we consider whether leaderboard size, that is, the number of other active participants on the leaderboard, is an important potential moderator of leaderboard impacts. Garcia et al. (2013) suggest that an important situational factor impacting comparison concerns and competitiveness is the number of competitors. On the one hand, increasing the number of active participants is likely to increase the likelihood of the positive dynamics that leaderboards introduce. Clearly, the mechanisms of competition, mutual accountability, and changes in perceived ability are nonexistent if there are no other active users on a leaderboard. More so, competitive motives may be stronger on larger leaderboards because ranking highly on larger leaderboards can be more motivational than dominating smaller leaderboards. That said, the effect of increasing leaderboard size is likely more nuanced. For instance, it is likely that some benefits of additional leaderboard participants are diminishing at the margin. Too many participants can make the leaderboard less effective because participants get lost in the crowd, weakening the positive impacts of competition or mutual accountability (Garcia and Tor 2009, Garcia et al. 2013). The diminishing marginal benefit of an additional leaderboard member implies nonlinearity in the benefit of more leaderboard members and may even lead to harmful effects of leaderboards if they become too large.

**3.3.2. Prior Activity Levels.** Second, we consider the physical activity level of an individual prior to leaderboard adoption.

*3.3.2.1. Competition.* If we consider only the role of competition (vis-à-vis prior activity levels), the expectation in the literature that highly active individuals should benefit disproportionately from leaderboards is most plausible (Patel et al. 2015, Wu et al. 2015, Shameli et al. 2017). Individuals with high activity levels prior to leaderboard adoption gain high utility from healthful activity and thus are likely to perform well on leaderboards. This positive performance on leaderboards can be motivational for them and encourage increases in future physical activity. The impact of competitive dynamics on relatively more sedentary individuals may be more nebulous. On the one hand, these individuals may benefit most from extrinsic motivators such as competition and ranking themselves against others. On the other hand, the value of leaderboards for such individuals may be limited by their lower intrinsic aptitude and motivation for physical activity. This leaves them prone to de-motivational impacts of lackluster performance on leaderboards.

*3.3.2.2. Accountability and Reference Points.* If we also consider theorized mechanisms related to social influence, the expectation ex ante is more uncertain. Because individuals on the low end of the physical activity distribution are more likely to have other leaderboard participants who are more active than they are, there is increased potential for the leaderboard to act as a tool that keeps them accountable; individuals who are more active than the focal user may be more credible in their attempts to hold the focal user accountable. More so, individuals at the lower end of the physical activity distribution are more likely to encounter other users who facilitate upward comparisons and positively impact their reference point for exercise and their perceived ability to engage in physical activity. In addition, individuals with low activity levels prior to leaderboard adoption may benefit most from leaderboards because they have more room for improvement and a higher need for external motivation. The dynamics around social influence are somewhat reversed for those who are highly active prior to leaderboard adoption. Following the same rationale, individuals who are already highly active may be less likely to join leaderboards where other users can hold them accountable (i.e., few others on their leaderboard can match their physical activity levels). In addition, these individuals are at elevated risk of leaderboards facilitating downward comparisons that negatively impact their exercise reference points. These comparisons can induce sluggishness if they highlight the focal user's disproportionate level of activity compared with others. Finally, highly active individuals may suffer from ceiling effects, that is, any extrinsic intervention is not likely to increase their willingness or ability to increase physical activity.

**3.3.3. Leaderboard Mechanisms, Prior Activity Levels, and Leaderboard Size.** The theorized effects of prior activity levels and leaderboard size can also intersect in ways that have implications for the diverse mechanisms through which leaderboards can impact behavior. First, our theorized mechanisms point to highly active individuals being most likely to be harmed by smaller leaderboards. Garcia et al. (2013) suggest that competitiveness emerges when there is a potential for comparisons, up or down, that credibly threaten the individual's rank. With smaller leaderboards (e.g., one other individual), these high achievers are less likely to interact with another individual who can credibly compete with them or hold them accountable (thus nullifying two key mechanisms for leaderboard value). At the same time, they are more likely to be presented with a salient individual who facilitates downward comparisons that negatively impact their exercise reference point, induce sluggishness, and diminish their physical activity levels. As leaderboard size increases, there is increased potential value for the highly active because the likelihood increases of at least one individual joining who can provide a credible threat to their rank, mutual accountability, and positive impacts on their reference point for exercise. Furthermore, it is plausible that individuals who are highly active are buoyed to perform even better when part of a relatively large leaderboard. This phenomenon would be akin to the idea in some sports of a "big match player," someone who performs above their average on big occasions and in front of big crowds.

In contrast, our theorized mechanisms have different implications for leaderboard size when individuals were sedentary prior to adoption. Unlike highly active individuals, these individuals can still benefit from adopting small leaderboards because they are still likely to encounter other users who are either at a comparable or a higher level of physical activity. Thus, even small leaderboards may often provide these individuals with an additional degree of accountability and the potential for positive impacts on their exercise reference points. Whether these individuals benefit from competition with small leaderboards is less certain, as they may still be dominated on small leaderboards, leading to de-motivational effects of competition. Increasing the size of the leaderboards for lower activity users may still provide some of the benefits described previously but it is likely that these benefits diminish faster for this group. Unlike for highly active individuals, benefits of mutual accountability may be reduced for these individuals as leaderboard size increases (via the "getting lost in the crowd" phenomenon described previously). More so, these users, who are at the lower end of the distribution of physical activity, may be more likely to get stuck toward the bottom of larger leaderboards and this may be more salient with more users participating. Overall, we conjecture that sedentary individuals can significantly benefit even when leaderboards are small. However, increases in leaderboard size may have diminishing marginal benefit for them.

## 4. Data and Model
### 4.1. Data
We use a unique panel data set comprised of 516 undergraduates at a U.S. university from October 2015 to September 2017.[11] This data set consists of granular wearable device data and periodic survey data. With respect to wearable device data, the students were offered Fitbit Charge HR devices, which were then used to record their physical activity. We access three types of Fitbit data: (i) step count, accessed on a daily basis; (ii) leaderboard data, which captures if a focal student has a leaderboard and, if so, the seven-day average step count of other leaderboard participants for the determination of participants' leaderboard rankings; and (iii) minute-by-minute heart rate data. Students synchronize their data with the Fitbit platform either through a dongle and a desktop application or a smartphone application. We implemented a client application that invoked the Fitbit application programming interface (API) to download the synchronized student activity data and store it locally in a secure database. The client application was a set of scripts that ran automatically every night. All study participants explicitly authorized our client application to allow access to their data via the Fitbit APIs.

Step measurements only occur if students wear their Fitbit devices regularly. We will use the term *compliance* to refer to the regularity with which students wear their Fitbit device. We calculate compliance from the heart rate data by assuming that a student is wearing their Fitbit during a particular minute of the day if the reported heart rate is nonzero. Students were paid $20 for maintaining at least 40% compliance and synchronizing their data regularly to Fitbit servers. Fitbit Charge HR could store up to seven days of data locally, so synchronizing beyond a seven-day interval would result in lost data and lower compliance.

Fitbit Charge HR's step measurements, which we use as the outcome in this study, are fairly accurate. Validation studies in laboratory and natural settings have found Fitbit Charge HR's mean absolute percent error (MAPE) for step count to be less than 10%, except for very light activity (Wahl et al. 2017, Bai et al. 2018). Bai et al. (2018) also found Fitbit Charge HR's heart rate measure to have an MAPE of ≈ 10%, although other studies have found mixed results. Even if the MAPE for heart rate were higher, our study is not likely to be negatively impacted. We use heart rate only for measuring compliance such that any nonzero heart rate measurement is construed as the device being used by the participant during that minute.

Participants were also asked to complete an intensive survey at the start of study and were further asked to take shorter surveys in six-month waves to refresh key measures. These surveys notably provided data on demographics (gender, religious affiliation, parent's income, etc.), psychological attributes using validated scales (personality, self-regulation), social interaction and ability (trust, anxiety, etc.), technology use (social media use, mobile app usage, etc.), and health state (body mass index, satisfaction with health, etc.). Although most students took the survey, there was some nonresponse as these surveys were not mandatory. On average, students completed three waves of survey data (approximately six months apart). We use these data in two ways. Primarily, we use relevant survey data to model the propensity for opting into a leaderboard and, in conjunction with advanced weighting approaches, construct a weighted sample that achieves covariate balance between leaderboard adopters and nonadopters. Secondarily, we use a subset of the survey data to generate controls that capture time-varying features of individuals that may relate to both leaderboard adoption and physical activity, and check the robustness of our main results.[12] Table 1 provides descriptive statistics about the outcome, treatment, and some demographic variables, whereas Online Appendix Table A.1 describes the relevant portions of the survey.

### 4.2. Model

The goal of our analysis is to estimate the effect of a user's leaderboard adoption on their physical activity as measured by steps walked, using nonexperimental data.[13] Thus, leaderboard adoption is the *treatment* in our observational study. In a randomized experiment, leaderboards could be randomly assigned to study participants, which would make the identification of treatment effect straightforward but would make the study treatment very different from naturally occurring leaderboards. In contrast, any Fitbit user in our study can opt into and construct their leaderboard, resulting in more natural leaderboards but making it more difficult to identify the treatment effect. The main empirical concern in identifying this treatment effect is the *confoundedness* of the leaderboard adoption with respect to users' physical activity as measured by their daily step count.

We use a DID research design as Fitbit users are observed over multiple time periods, with roughly half of the users adopting leaderboards and the other half remaining untreated. Although a DID design controls for any time-invariant user characteristics and common shocks, it requires the identifying assumption that any uncontrolled time-varying user characteristics exhibit a common trend across the treated and untreated individuals. Under this identifying assumption, we can estimate the effect of leaderboards on steps walked for the Fitbit users who have adopted leaderboards. Our model specification is given here and its explanation follows:

$$Steps_{it} = \beta_0 + \beta_1(Leaderboard_{it}) + \theta_i + \lambda_t \\ + \gamma_i \times t + \phi_i \times t^2 + \epsilon_{it}. \quad (1)$$

Although we observe physical activity data on a daily basis, the leaderboard data are obtained weekly. Hence, our unit of analysis is student-week. $Steps_{it}$ is the average number of steps walked daily by user $i$ in week $t$, and $Leaderboard_{it}$ is a binary indicator for whether a user $i$ adopted a leaderboard in week $t$. As stated earlier, the leaderboard adoption is generally "sticky," as Fitbit makes it difficult to de-adopt. We include individual fixed effects ($\theta_i$) to account for time-invariant differences between individuals and time-fixed effects ($\lambda_t$) to account for any common shocks in our data. Together, these two-way fixed effects would enable the identification of the treatment effect in the absence of any differential trends across the treated and untreated individuals. Admittedly, the common trends assumption is very strong, but one way to make it more plausible is to explicitly control for individual-specific linear time trend ($\gamma_i \times t$) and individual-specific quadratic time trend ($\phi_i \times t^2$). We can then estimate our model under the weaker assumption that the treatment assignment is ignorable after controlling for the two-way fixed effects and the additional individual-specific time trends (Xu 2017).[14] In Section 5.1, we will further explore the issue of time trends across Fitbit users with and without leaderboards. For inference, all our analyses use cluster-robust variance-covariance estimators (VCE), clustered at the student level, which adjust for heteroskedasticity and serial correlation.

**Table 1.** Descriptive Statistics

| Variable | Description | Mean | Standard deviation | Minimum | Maximum |
|---|---|---|---|---|---|
| Steps | Number of steps walked daily | 10,625.18 | 4,247.53 | 0 | 37,835 |
| Leaderboard | An indicator if an individual has adopted a leaderboard | 0.47 | 0.50 | 0 | 1 |
| Age | Age at the start of the study | 17.94 | 0.72 | 17 | 26 |
| Body mass index | Body mass index at the start of the study | 22.74 | 3.02 | 16 | 38 |
| Female | An indicator for whether the individual is a female | 0.50 | 0.50 | 0 | 1 |
| Leaderboard size | The number of users on the leaderboard (excluding the focal users) | 4.78 | 4.46 | 1 | 25 |
| Leaderboard size (active) | The number of users on the leaderboard that have a nonzero step count (excluding the focal users) | 2.32 | 2.64 | 0 | 17 |

# 5. Estimation and Robustness of the Main Effects of Leaderboards

In our main analysis, we estimate variants of Specification (1), which is a DID specification with flexible user-specific time trends. Table 2, columns 1 and 2, presents the estimation results for Specification (1). The first column presents results for a specification that includes individual-specific linear time trends only, whereas the second column additionally includes individual-specific quadratic time trends. In both columns, we find a significant ($p < 0.05$) and meaningful leaderboard effect of 338–370 steps daily. Column 2 is our preferred model as it includes more flexible time trends. This model suggest that the students who adopted leaderboards have a daily increase of 370 steps, equivalent to a 3.5% increase in physical activity on the average daily step count of 10,268. These initial results suggest some support for a main effect of leaderboard adoption on physical activity. In the Online Appendix, we extend this analysis to add time-varying survey variables as controls in Specification (1). The estimated effects with these additional controls have higher magnitudes, which increases the plausibility of the main results. However, a number of concerns commonly arise with analyses using observational data. In the remainder of this section, we discuss the robustness of our main results.

## 5.1. Probing the Common Trends Assumption

Identification of the treatment effect with a DID design crucially depends on the common trends assumption. As mentioned earlier, one way to weaken this assumption is to control for individual-specific linear and quadratic time trends, which we have incorporated in our model estimation. In this section, we will further probe the plausibility of assuming that no unobserved time-varying covariates may be confounding our analysis (i.e., the common trends assumption).

### 5.1.1. Inverse Probability of Treatment Weighting.
A common concern in a DID design is whether the treated and control subjects are similar in their baseline characteristics such that the treated and control subjects plausibly follow common trends. As mentioned earlier, we collected a rich set of baseline characteristics of study users using a survey instrument. Although the initial covariate balance did not cause excessive concern, we use the inverse probability of treatment weighting (IPTW) method to further improve the covariate balance of our sample. To estimate the propensity for leaderboard adoption, we use the Toolkit for Weighting and Analysis of Nonequivalent Groups (TWANG), which implements a generalized boosted regression model (GBM). The propensity score estimated by TWANG optimizes covariate balance across leaderboard adopters and nonadopters. We observe substantive improvement in the postweighting covariate balance such that the observed absolute standardized mean difference, $SMD \leq 0.2$, is better than the accepted threshold of 0.25.[15] Table 2, column 3, presents the main analysis using the IPTW sample. The sample size is slightly smaller (cf. column 2) as a few students did not participate in the initial study survey. Comparing with the main result (column 2), we find the effect sizes to be very similar—370 versus 343 steps. This stability of effect size boosts our confidence in the main results.

### 5.1.2. Pretreatment Period Placebo Treatments.
Given that we have multiple pretreatment periods for most users in our sample, we can probe the plausibility of the common trends assumption by creating placebo treatments in

**Table 2.** Fitness Activity and Leaderboard Participation

| | (1) Steps b/se | (2) Steps b/se | (3) Steps b/se | (4) Steps b/se | (5) Steps b/se | (6) Steps b/se | (7) Steps b/se |
|---|---|---|---|---|---|---|---|
| *Leaderboard* | 338.38** (171.40) | 370.46** (170.66) | 343.02** (170.98) | | | 383.20** (190.75) | 397.75** (174.05) |
| *Placebo Leaderboard 4-Weeks* | | | | −3.62 (260.57) | | | |
| *Actual Leaderboard 4-Weeks* | | | | | 419.78[†] (271.53) | | |
| *Leaderboard × Inviter* | | | | | | −98.92 (387.32) | |
| Individual fixed effects | Yes | Yes | Yes | Yes | Yes | Yes | Yes |
| Week fixed effects | Yes | Yes | Yes | Yes | Yes | Yes | Yes |
| Individual linear trends | Yes | Yes | Yes | Yes | Yes | Yes | Yes |
| Individual quadratic trends | No | Yes | Yes | Yes | Yes | Yes | Yes |
| IPTW | No | No | Yes | No | No | No | No |
| Observations | 27,758 | 27,758 | 27,409 | 14,746 | 15,742 | 27,758 | 27,358 |
| Individuals | 516 | 516 | 501 | 516 | 516 | 516 | 503 |
| Adjusted $R^2$ | 0.3 | 0.33 | 0.34 | 0.34 | 0.35 | 0.33 | 0.33 |
| VCE | Robust | Robust | Robust | Robust | Robust | Robust | Robust |

*Notes.* Column 7 (cf. column 2) excludes users who hide themselves. Please see Section 5.4 for details.
[†]$p < 0.125$; *$p < 0.10$; **$p < 0.05$.

the pretreatment data alone, that is, by dropping the posttreatment data and using only the pretreatment data for this analysis. A failure to reject the null effect for the placebo treatment would provide support for the common trends assumption.[16] In our study, users opt into the treatment in different periods. Moreover, our primary concern is the presence of some unobserved time-varying factor (e.g., spurts in motivation) that affected the adopters in the periods closely preceding the treatment. Hence, we implemented our placebo treatment in the preceding month prior to the actual treatment and estimated the model in Equation (1) on the altered data. Table 2, column 4, presents the estimated effect of the placebo treatment. This estimated effect is small in magnitude, opposite in sign, and statistically insignificant. However, if we include four weeks of actual treatment period in our sample, the estimated effect is $\approx 420$ steps ($p = 0.12$) as presented in Table 2, column 5. Thus, for comparable time periods, the placebo effect is null, whereas the actual treatment effect is comparable to our estimated main effect. This null effect in the pretreatment period enhances the plausibility of our common trends assumption.

*5.1.2.1. Leads-Lags Model.* The placebo treatment effect can be further broken into weekly placebo effects in the pretreatment period and the actual effect in the posttreatment period using the full data set and a leads-lags specification:

$$Steps_{it} = \beta_0 + \sum_{\tau=-9}^{9} \beta_\tau L_{i(t+\tau)} + \theta_i + \lambda_t + \gamma_i \times t + \phi_i \times t^2 + \epsilon_{it}. \quad (2)$$

The dummy variables $L_{i(t+\tau)}$ denote the time from adoption; for example, $L_{i(t+\tau)}$ would be one for individual "$i$" in time period "$t$" if this time period is $\tau$ weeks from adoption, where $\tau = -10, \ldots, 9$. If we observe more data for an individual, we collapse it into the extreme periods.[17] We set $L_{i(t-10)}$ as the baseline period and exclude it from Equation (2) to avoid the "dummy variable trap." Figure 2 (top left panel) plots point estimates and confidence intervals for $\beta_\tau$ coefficients against the time from adoption. We find a null effect in the pretreatment period. In contrast, the effect is positive and statistically significant in the adoption period (i.e., period 0). The posttreatment coefficients remain positive but decrease in magnitude and lose significance in the later periods. We will explain the reason for this decline in Section 6.1.1. A potential issue with this analysis is that the coefficients are trending upward from period $-4$ to $-1$. However, this trend is not a cause of concern for several reasons: first, the estimates for periods $-4$ to $-1$ are not statistically significant even at the 75% level (the lowest $p$ value is 0.29). Second, the estimates for $\beta_\tau$ in the posttreatment period stay positive (and larger than any preperiod estimate), whereas the estimated $\beta_\tau$ in the preperiod oscillate around the zero line. In particular, the change in coefficient estimates from period $-9$ to $-6$ is roughly the same as the change from $-4$ to $-1$, with a sharp drop to a very small negative value in period $-5$. Thus, extrapolating this historical pattern beyond $-1$ would plausibly suggest a regression back to an almost zero value as in period $-5$, but the adoption breaks that trend such that we see a large significant effect in the adoption period and beyond. Finally, the other tests such as the placebo test presented earlier in this section also argue against the presence of any pretrend in the month preceding the adoption. These robustness checks argue against the presence of any other unmeasured changes that affect leaderboard adopters in the periods closely preceding leaderboard adoption, thus enhancing the plausibility of the common trends assumption.

### 5.2. Robustness Check for Leaderboard Initiation
As an additional robustness check, we also considered leaderboard initiation as it may be a proxy for confounded leaderboard adoption. Specifically, if the focal user is the primary inviter to the leaderboard, this leaderboard may be more likely to be driven by unobserved motivation changes. For the purpose of this analysis, we consider focal users to be of the "inviter-type" if they initiate most, not necessarily all, of the invitations to other users on their leaderboard. Although we do not have access to direct measures of who initiated a leaderboard, we construct a proxy variable that we argue identifies users who are more likely to be inviter types. We leverage two aspects of leaderboard creation to construct this proxy variable. First, per the discussion in Section 2.1, Fitbit does not use a leaderboard that is defined centrally as a group of individuals that others can join or leave. Rather, each leaderboard is owned by the user and each user pair must agree to share their step information for them to be joined on their individual leaderboards. In addition, Fitbit does not advertise to the user's friends that they have joined the platform.

Based on these aspects of Fitbit leaderboards, we designated *InviterLB* using two criteria: (i) whether the leaderboard had three or more individuals when it was first adopted and (ii) whether the leaderboard was such that the other users (excluding the focal users) had been on the Fitbit platform for longer than 90 days. The first criterion is useful because the size of the leaderboard at leaderboard initiation can be indicative of the likelihood of initiation by the focal user. If there are two people when the leaderboard is started, it is unclear who initiated. However, if three people (or more) are on the leaderboard at initiation, a leaderboard fully initiated by others would require that two other users actively searched and invited the focal user in the same week and that the user accepted both invitations. However, this criterion may still include mixed leaderboards that were only partially

**Figure 2.** (Color online) Leaderboard Coefficients by Weeks from Leaderboard Adoption (by Prior Activity Levels)

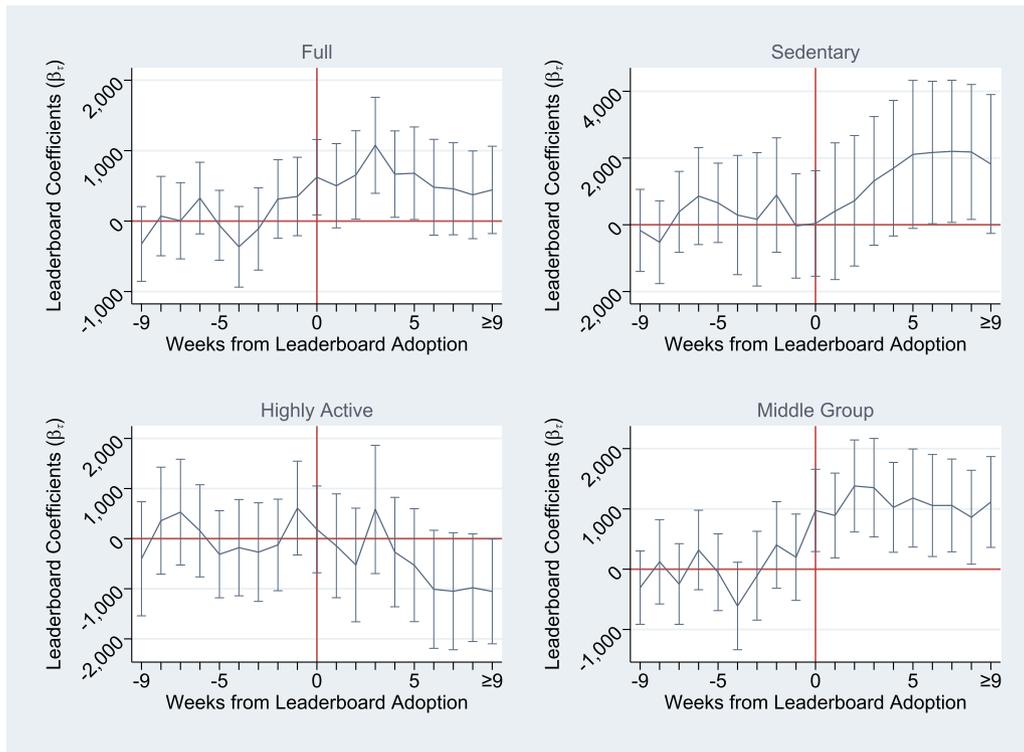

*Note.* (a) Please see Equation (2) for charts' specification, (b) 90% confidence intervals, (c) vertical axes use different scales.

initiated by the focal users (e.g., another user initiates but the focal user invites the third person). Thus, we add the second criterion that the other users on the leaderboards have been on the platform for more than 90 days. The rationale behind this criterion is that users on the platform for longer periods of time are more settled on the platform and less likely to be actively scouring the platform for new connections. The 90-day threshold was chosen based on data suggesting that Fitbit abandonment happens in the first few months of adoption.[18] Among the leaderboard adopters, 15.3% met these criteria. In Table 2, column 6, we add an interaction term between leaderboard and an indicator for *InviterLB* and identify a negative coefficient that is close to zero and insignificant ($p = 0.8$). This result suggests that users who are more likely to have initiated the leaderboard do not see different treatment effects and is further evidence that time-varying changes in motivation are unlikely to be confounding our results.

### 5.3. Fitbit Compliance

Step measurement through Fitbit only occurs if the participants wear their devices regularly. We will use the term *compliance* to refer to the regularity with which a participant wears the Fitbit device. In this subsection, we will probe two compliance-related concerns that may cast doubt on the earlier analyses if left unaddressed. Fortunately, our data include compliance data at a very granular level, which allows us to construct the participant's compliance measure, *percent compliance*, at the daily and weekly level, and the participant's *mean compliance* for the study duration. We will exploit these data to probe compliance-related concerns.

**5.3.1. Do Leaderboards Increase Compliance?** The first concern is the possibility that rather than increasing steps, leaderboard adoption increases compliance, which may lead us to observe higher step count purely because of better measurement. To address this concern, we estimated main effects models similar to Equation (1) and the leads-lags models similar to Equation (2) but with daily percentage compliance as the dependent variable and student-day as the unit of analysis. We estimate this model for a number of samples—the entire sample and the subsamples at various mean compliance levels (ranging from 60% to 95%). The main effect model estimates are statistically insignificant and have small magnitudes, ranging from –1.97% to 2.01%. In addition, the leads-lag model's coefficient plots do not show any sharp increase at or after adoption (see the Online Appendix, Section E.1). These results suggest a null effect of leaderboard adoption on compliance.

**5.3.2. Are Leaderboard Effects Discernible at Higher Compliance Levels?** The second issue is that some of the participants may have lower compliance and the full sample estimate includes these participants too.

 

Regarding this concern, our empirical analysis would be more convincing if the leaderboard's effect on participant activity was clearly discernible for participants with high compliance levels. Thus, we estimate impact of leaderboards for participants with high levels of compliance and find these effects to range from 408 to 598 steps (see the Online Appendix, Table E.6). The persistence of leaderboard effects at higher levels of compliance supports the claims from our main results.

### 5.4. Fitbit Attrition, Leaderboard De-Adoption, and Additional Robustness Checks

Related to the challenge of compliance, we also consider the role of attrition from the sample due to Fitbit abandonment, which has generally been noted in the popular press for health wearables.[18] If sample attrition is related to leaderboard adoption, it may introduce bias in our analysis. For example, lower performers may abandon their Fitbit device after joining leaderboards because it reveals to them that they are less active than their peers. We examine this concern extensively and identify no relationship between leaderboard adoption and sample attrition for lower performers, and no differences in physical activity and similar leaderboard effects for those who eventually leave the sample compared with those who report data throughout (see the Online Appendix, Section F). We also consider whether individuals who eventually hide themselves from the leaderboard (the main mechanism for de-adoption) impact our results. As we mentioned previously, this was rare for leaderboard adopters (approximately 5%), and excluding these individuals results in consistent estimates of leaderboard effects (see Table 2, column 7).

#### 5.4.1. Outliers and Falsification with Negative Control Treatments.
We also evaluate the potential for a particular individual (or time period) in the data to be an outlier driving our results. Specifically, we systematically "leave out one" individual (or time period) and re-estimate our model (see the Online Appendix, Section G). We find consistent treatment effects of leaderboards that are always statistically significant, suggesting minimal risk from outliers in the data. Furthermore, we constructed a negative control treatment (NCT), as the focal user's leaderboard with no other active users. Such leaderboards exist because other users may accept a request to connect but then become inactive on the platform and thus neither provide competition nor reference points (see the Online Appendix, Figure D.II and associated discussion). Thus, the absence of any other active users of such leaderboards should result in no effect on the user's physical activity. Indeed, we find a null effect of such leaderboards on steps (see the Online Appendix, Section D). This falsification test with an NCT strengthens the plausibility of the common trends assumption.

## 6. Heterogeneous Effect of Leaderboards

In this section, we evaluate the potential for heterogeneous effects of leaderboards on physical activity focusing on leaderboard rank, leaderboard size, and prior activity level. The evaluation of heterogeneous effects of leaderboards is useful because it can offer additional insights into the role of competition and social influence in generating leaderboard value.

We start by evaluating the impact of ranking first on the leaderboard in the prior period on the activity levels of individuals in the subsequent period ($FirstOnLB$).[19] Next, we evaluate whether the number of active participants on a leaderboard (excluding the focal user) modifies the benefit to individuals who adopt leaderboards ($LBActiveUsers$) and whether this impact is nonlinear, by including the square of ($LBActiveUsers_{it}$).[20] Finally, we consider the interaction of rank and leaderboard size. Equation (3) provides the specification for this model. To evaluate heterogeneous impacts by prior activity levels, we also estimate this specification stratified by preleaderboard activity levels.

$$\begin{aligned} Steps_{it} = {} & \beta_0 + \beta_1(FirstOnLB_{it-1}) + \beta_2(LeaderBoard_{it}) \\ & + \beta_3(LBActiveUsers_{it}) + \beta_4(LBActiveUsers_{it})^2 \\ & + \beta_5(FirstOnLB_{it-1} * LBActiveUsers_{it}) \\ & + \beta_6(FirstOnLB_{it-1} * (LBActiveUsers_{it})^2) \\ & + \theta_i + \lambda_t + \gamma_i \times t + \phi_i \times t^2 + \epsilon_{it} \end{aligned} \qquad (3)$$

#### 6.0.1. Impact of Prior Week's Rank and Leaderboard Size.
We find a substantive impact of prior week's rank on the impact of leaderboards in subsequent periods. Those who were in first place on their leaderboard in one week, walked 578 steps more a day the following week (Table 3, column 1). We also find a positive and significant coefficient on *LBActiveUsers* (Table 3, column 2), suggesting that an additional person on a leaderboard increases the effect of that leaderboard by 165 steps ($p < 0.05$). Moreover, we evaluate whether there are diminishing benefits from additional active users on a leaderboard by adding a quadratic term to our estimation (Table 3, column 3). A negative coefficient on the quadratic term ($\approx -21$) suggests that the effect peaks at eight active members, after which the marginal benefit of an additional member is diminishing.

In Table 3, columns 4 and 5, we also explore the intersectional impact of leaderboard size and prior performance by including $FirstOnLB$ and $FirstOnLB \times (LBActiveUsers)^k$, $k = 1, 2$. We find that the motivational effects of succeeding in leaderboard competition increase with the size of the leaderboard. More so, we continue to identify a positive impact of increased leaderboard size, suggesting positive impacts of larger leaderboards when an individual is not first. Finally, we consider whether the nonlinear impacts of leaderboard size extend to the interaction with prior week performance

**Table 3.** Heterogeneous Effect by Leaderboard Size and User Rank

|  | (1) Steps b/se | (2) Steps b/se | (3) Steps b/se | (4) Steps b/se | (5) Steps b/se |
| --- | --- | --- | --- | --- | --- |
| *Leaderboard* | 219.28 (170.84) | 108.99 (183.39) | −61.66 (187.25) | −29.60 (184.02) | −194.81 (188.89) |
| *FirstOnLB* | 577.98** (94.80) |  |  | 371.70** (111.31) | 377.58** (130.19) |
| *LBActiveUsers* |  | 164.46** (31.50) | 334.31** (57.55) | 155.28** (32.22) | 315.90** (56.89) |
| *(LBActiveUsers)$^2$* |  |  | −21.36** (6.72) |  | −20.00** (6.47) |
| *FirstOnLB × LBActiveUsers* |  |  |  | 146.23** (43.49) | 168.20* (99.07) |
| *FirstOnLB × (LBActiveUsers)$^2$* |  |  |  |  | −5.91 (10.96) |
| Individual fixed effects | Yes | Yes | Yes | Yes | Yes |
| Week fixed effects | Yes | Yes | Yes | Yes | Yes |
| Individual linear trends | Yes | Yes | Yes | Yes | Yes |
| Individual quadratic trends | Yes | Yes | Yes | Yes | Yes |
| Observations | 27,758 | 27,758 | 27,758 | 27,758 | 27,758 |
| Individuals | 516 | 516 | 516 | 516 | 516 |
| Adjusted $R^2$ | 0.33 | 0.33 | 0.33 | 0.33 | 0.34 |
| VCE | Robust | Robust | Robust | Robust | Robust |

*$p < 0.10$; **$p < 0.05$.

and find no evidence of diminishing impacts (coefficient for *FirstonLB × (LBActiveUsers)$^2$* is near 0 and insignificant). Overall, we find that leaderboard benefit increases with leaderboard size (although this benefit is diminishing at the margin), and that ranking highly on a leaderboard is more motivational on larger leaderboards.

**6.0.2. Implications of Findings.** The positive impact of prior rank and the increased impact of ranking first on larger leaderboards point to an important role of competitive dynamics in generating leaderboard value. In addition, the impact of leaderboard size above and beyond the impact of rank suggests that social influence mechanisms also play an important role in observed leaderboard benefits. However, an insignificant effect of small leaderboard when the individual is not ranked first and diminishing marginal benefit of increasing leaderboard size suggest some nuance around how social influence mechanisms drive leaderboard benefits. We explore this nuance further by evaluating how prior activity levels (which have implications for how other users on leaderboards exert social influence) moderate leaderboard impact.

### 6.1. Heterogeneity by Prior Activity Levels

To evaluate potential heterogeneity in leaderboard benefit by prior activity, we estimate Specification (1) on samples stratified by preleaderboard activity levels. Specifically, we stratify our sample into two groups based on their preleaderboard activity levels: the top quartile by daily step count comprise the highly active group whereas the bottom quartile by daily step count comprise the sedentary group.[21] The differences in step count between these two groups were meaningful in terms of magnitude and were statistically significant (13,000 versus 8,000, $p < 0.01$)—see the Online Appendix, Table H.9, for summary statistics on these groups.

Before examining the impact of leaderboards on physical activity levels, we evaluated the correlation between activity levels and relevant preadoption survey measures (see the Online Appendix, Table H.10). We found correlations consistent with our expectations of the key differences between highly active and sedentary individuals. Sedentary individuals reported lower levels of self-efficacy and self-regulation for exercise and reported being more likely to exercise alone. In addition, sedentary individuals reported higher levels of anxiety and depression and lower levels of trust. These correlations suggest that sedentary individuals may need these interventions more than highly active individuals but that they could also be prone to de-motivational impacts if these leaderboards exacerbate mental health barriers to improving health (e.g., increase their anxiety).

Turning to the examination of the effect on steps, we find stark differences in the effect of leaderboards on the highly active group vs. the sedentary group. For sedentary users, the adoption of leaderboards has large and significant impacts on their daily step counts (1,365, $p < 0.05$; see Table 4, column 1). In contrast, we find that the highly active group, instead of benefiting from leaderboards, experienced a significant decrease in their daily physical activity after leaderboard adoption (–631, $p < 0.05$; see Table 4, column 2). For the sake of completion, we also estimate the effect for the middle group, i.e.,

**Table 4.** Heterogeneous Effect by Leaderboard Size, Rank, and Prior Activity

|  | (1) Steps b/se | (2) Steps b/se | (3) Steps b/se | (4) Steps b/se | (5) Steps b/se | (6) Steps b/se | (7) Steps b/se |
|---|---|---|---|---|---|---|---|
| *Leaderboard* | 1,365.44** | −630.63** | 858.60** | 1,262.11** | −817.47** | 1,016.43** | −1,215.74** |
|  | (496.72) | (292.10) | (213.46) | (484.75) | (304.86) | (486.63) | (331.38) |
| *FirstOnLB* |  |  |  | 745.74* | 522.47** | −49.60 | 257.18 |
|  |  |  |  | (422.44) | (157.44) | (343.07) | (198.09) |
| *LBActiveUsers* |  |  |  |  |  | 214.90** | 208.89** |
|  |  |  |  |  |  | (46.52) | (61.16) |
| *LBActiveUsers × FirstOnLB* |  |  |  |  |  | 462.99** | 149.33** |
|  |  |  |  |  |  | (87.96) | (55.44) |
| Sample | S | HA | Mid | S | HA | S | HA |
| Individual fixed effects | Yes | Yes | Yes | Yes | Yes | Yes | Yes |
| Week fixed effects | Yes | Yes | Yes | Yes | Yes | Yes | Yes |
| Individual linear trends | Yes | Yes | Yes | Yes | Yes | Yes | Yes |
| Individual quadratic trends | Yes | Yes | Yes | Yes | Yes | Yes | Yes |
| Observations | 5,629 | 7,836 | 14,293 | 5,629 | 7,836 | 5,629 | 7,836 |
| Individuals | 129 | 129 | 258 | 129 | 129 | 129 | 129 |
| Adjusted $R^2$ | 0.38 | 0.34 | 0.32 | 0.38 | 0.34 | 0.38 | 0.34 |
| VCE | Robust | Robust | Robust | Robust | Robust | Robust | Robust |

*Note.* S, sedentary; Mid, mid 50 percentile; HA, highly active.
  *$p < 0.10$; **$p < 0.05$.

individuals whose activity levels were between the 25th and 75th percentile. We find that the middle group benefited from leaderboards, but the benefit was less than for those who were in the bottom quartile (859, $p < .05$; see Table 4, column 3). Overall, these results suggest significant heterogeneity in the effect of leaderboard based on prior activity levels, with particularly stark, negative effects for those who were previously highly active.[22]

**6.1.1. Leads-Lags Model by Prior Activity.** To examine the presence of any trends before leaderboard adoption, we plotted the coefficients from a leads-lags model for the full sample and subsamples by prior activity levels in Figure 2 (see Section 5.1.2 for details about the specification). The first point to note is that the lead coefficients are relatively small in magnitude and statistically insignificant, which increases the plausibility of the parallel trends assumption. Second, the lag coefficients shift to larger magnitudes with the effect persisting for more than two months after adoption. Third, there is some attenuation in the effect for the full sample in later time periods, which can be plausibly explained by the opposite direction of effect within the sedentary and middle groups versus the highly active subsample.

**6.1.2. Implication of Findings.** The divergence of leaderboard effects for sedentary versus highly active individuals substantiates our conjecture that leaderboards can introduce both motivational and de-motivational dynamics with respect to physical activity. Specifically, these results are suggestive evidence of different impacts on reference points for sedentary versus highly active users and differences in the potential of leaderboards to provide accountability for lapses in physical activity. We explore this difference and its implications for mechanisms underlying leaderboard value further by evaluating the impact of rank and leaderboard size on the physical activity of sedentary versus highly active users in the next section.

### 6.2. Interaction of Leaderboard Size, Rank, and Prior Activity Levels
Last, we consider the intersection of all the prior factors in an effort to understand some of the heterogeneity in leaderboard benefit for sedentary versus active users.

**6.2.1. Leaderboard Size and Rank.** We start by evaluating the motivational impact of leaderboard rank and size separately by prior activity levels. We find that both previously sedentary and highly active individuals see substantial increases in physical activity during the week after they ranked first on their leaderboard (Table 4, columns 4 and 5). For sedentary individuals, being first in the prior week unlocks even more value for them and increases the treatment effect for leaderboards by 746 steps to nearly 2,000 steps. Highly active individuals see slightly less value (522 steps versus 746 steps) from being first in the prior week but this benefit counteracts part of the negative main effect they observe. Extending this analysis to also include leaderboard size and the interaction of leaderboard size and prior week's performance reveals further richness in these results. In column 6, we observe that previously sedentary individuals still see substantial benefit when leaderboards are small and when they are not first (1,016 steps, $p < 0.1$) and that this benefit increases further when they rank first and leaderboards are larger. In column 7, we observe that previously active individuals are harmed when they are on small leaderboards; these users only start to see positive impacts from leaderboards if they rank first on leaderboards with more than four individuals or if they are on relatively large leaderboards (more than six active users).

The finding that sedentary individuals observe leaderboard benefit despite unsuccessfully competing on small leaderboards suggests that these individuals accrue significant benefit from noncompetition mechanisms (i.e., social influence). In contrast, the negative impact of the same type of leaderboard for highly active individuals suggests that social influence is harming these individuals, or that its benefit is not sufficient to overcome any de-motivational effects of not ranking first. We probe this conclusion further using two additional analyses that leverage certain leaderboard instances which could be telling of the impact of these noncompetition mechanisms on physical activity. The first analysis seeks to identify instances of leaderboards where competitive dynamics are arguably weakened but the potential for mutual accountability and changes in reference points is still present. Specifically, we create *NoCompetitionLB*, which is an indicator of leaderboard instances where the focal user is sandwiched between two other users such that there is not a credible threat to their rank from either user (see the Online Appendix, Section H.3 for details). Table 5, columns 1 and 2, shows a positive and significant impact (534 steps, $p < 0.05$) of these types of leaderboards for previously sedentary users but a small and insignificant impact of these same types of leaderboards for highly active users ($-39$ steps, $p = 0.9$). We examine this conjecture further by identifying instances of leaderboards at the other end of the competition spectrum. Specifically, we create *HighCompetitionLB* as an indicator for leaderboard instances where the focal user is regularly being displaced and then reclaiming the top spot on the leaderboard (see the Online Appendix, Section H.3 for details). Table 5, column 3, shows that sedentary users continue to perform well on leaderboards without high competition intensity, with no statistically significant benefit to them of being on a highly competitive leaderboard. In contrast, highly active users have large negative effects when they are on leaderboards without intense competition, but there is a statistically significant offsetting effect of being on leaderboards with high competition (Table 5, column 4).

**6.2.2. Implication of Findings.** Although only suggestive evidence, these results lend some credence to the notion that previously sedentary individuals are benefiting from mutual accountability and positive impacts on their exercise reference point and do not need competition to benefit. In contrast, highly active individuals seemed to be harmed by the noncompetition mechanisms, but this harm can be offset if leaderboards are sufficiently competitive.

**6.2.3. Nonlinear Impacts of Leaderboard Size.** Finally, we consider whether nonlinear impacts of leaderboard size are similar based on prior activity level. In Table 5, columns 5 and 6, we find that, although both groups have diminishing returns from additional users, the negative coefficient on the quadratic term ($ActiveLB^2$) for the sedentary group is thrice that for the highly active group. These results suggest that for sedentary

**Table 5.** Further Heterogeneous Effect Analysis

| | (1) Steps b/se | (2) Steps b/se | (3) Steps b/se | (4) Steps b/se | (5) Steps b/se | (6) Steps b/se |
|---|---|---|---|---|---|---|
| *NoCompetitionLB* | 533.74** (261.125) | −38.98 (283.539) | | | | |
| *Leaderboard (LB)* | | | 1,283.58** (560.852) | −1,007.23** (343.563) | 767.49 (516.114) | −1,268.45** (316.771) |
| *LB × High Competition* | | | 660.56 (1,234.148) | 1,021.99* (611.764) | | |
| *LBActiveUsers* | | | | | 610.05** (233.570) | 400.18** (89.988) |
| *(LBActiveUsers)$^2$* | | | | | −57.74* (31.635) | −19.86** (7.935) |
| Sample | S | HA | S | HA | S | HA |
| Individual fixed effects | Yes | Yes | Yes | Yes | Yes | Yes |
| Week fixed effects | Yes | Yes | Yes | Yes | Yes | Yes |
| Individual linear trends | Yes | Yes | Yes | Yes | Yes | Yes |
| Individual quadratic trends | Yes | Yes | Yes | Yes | Yes | Yes |
| Observations | 5,500 | 7,707 | 5,629 | 7,836 | 5,629 | 7,836 |
| Individuals | 124 | 129 | 129 | 129 | 129 | 129 |
| Adjusted $R^2$ | 0.38 | 0.33 | 0.38 | 0.34 | 0.38 | 0.34 |
| VCE | Robust | Robust | Robust | Robust | Robust | Robust |

*Note.* S, sedentary; HA, highly active.
\*$p < 0.10$; \*\*$p < 0.05$.

**Figure 3.** (Color online) Heterogeneous Effect by Active Users and Prior Activity

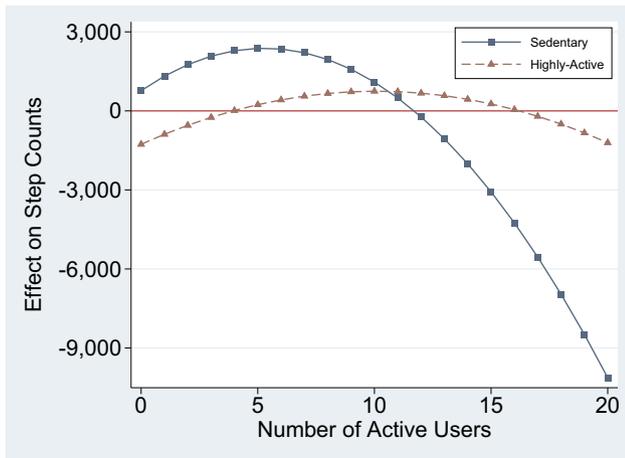

users, the benefits of an additional active leaderboard member diminish much faster (after 5 users) than they do for highly active users (after 11 users; Figure 3).

**6.2.4. Implications of Findings.** The smaller optimal size for sedentary individuals suggest that they accrue benefits (e.g., positive impacts on their exercise reference points) even when leaderboards are small. However, benefits diminish as leaderboards become larger, and leaderboards can even be harmful if they become excessively large (leaderboard sizes that become harmful to sedentary users were uncommon in our data). In contrast, highly active individuals seem to be demotivated by leaderboards with too few individuals and only start to benefit on larger, more competitive leaderboards. These dynamics for highly active individuals reinforce the notion that these users require large leaderboards to derive benefit.

### 6.3. Summary of Findings from Heterogeneous Effect Analysis

Ex ante, we theorized that competition and social influence are key mechanisms underlying leaderboard effects but that these mechanisms may introduce both motivational and de-motivational effects of leaderboards. The heterogeneous effects we identify in this section point to the importance of these mechanisms as well as the potential for nuance in their effects. First, we find robust positive impacts of ranking first on a leaderboard, suggesting that successfully competing on leaderboards improves motivation for most users. Interestingly, and contrary to our theoretical conjecture, the benefits of competition hold even for sedentary users: they accrue positive effects from ranking first and are not harmed from leaderboards when they do not rank first.

We attribute the robust benefit of leaderboards for sedentary users to the positive impacts of leaderboards on their exercise reference points and the likelihood of being held accountable by other users (i.e., social influence). However, our results also suggest that social influence enabled by leaderboards can have negative impacts on motivation for some users (e.g., negative impacts on exercise reference points for highly active users). Interestingly, these harms for highly active individuals are attenuated when leaderboards are highly competitive. The nuanced impact of social influence is further demonstrated by the impacts of leaderboard size on physical activity. While larger leaderboards generally increase physical activity, sedentary users see diminishing value from larger leaderboards. This result is in line with our conjecture that social influence effects may diminish for sedentary users if they get "lost in the crowd" of larger leaderboards. In contrast, highly active individuals thrive in large leaderboards, substantiating our conjecture that highly active individuals become more likely to show positive impacts of competition and social influence as leaderboard size increases.

## 7. Conclusions and Discussion

The rapid and increasingly broad adoption of health wearables coupled with the gamification services built on top of them provides a potentially powerful vehicle for improving health behaviors at scale. Our results lend credence to this potential value, particularly when considered over time. In our data, the average user participated in a leaderboard for 237 days (conditional on participating in a leaderboard). Thus, sedentary users who participated in a leaderboard for at least 237 days took more than 300,000 additional steps (with a conservative estimate of 1,300 additional daily steps). To put this in context, the aggregated benefit of leaderboards (for these participants) amounts to 150 miles of distance (at 2,000 steps a mile). Importantly, the benefit is not homogeneous and there is a decrease in daily steps of nearly equal measure for those who, prior to adoption, were highly active. However, this health harm to the highly active subgroup may not be symmetrical to the gain for those who were previously sedentary, as this group remains very active in absolute terms. We also identify additional heterogeneity in benefit based on the number of other active participants and leaderboard rank.

This research has some important limitations. First, we use secondary data in which individuals organically choose to adopt leaderboards, rather than being randomly assigned to the treatment. Although we put in significant effort to address potential confounding factors for our analysis, only a large-scale randomized control trial can provide a theoretical guarantee that the treatment is unconfounded. Additionally, we were missing variables in our data set which can be particularly informative of leaderboard mechanisms and the heterogeneous value they generate. Specifically, we do not have deterministic measures of whether the focal user initiated the leaderboard or whether it was initiated by another user. A more comprehensive measure of leaderboard initiation could have provided important insights into how different types of users are initiating leaderboards, whether initiating a leaderboard impacts competitive dynamics, and whether individuals are selecting into leaderboards of value to them or if others are prompting them to do so. We hope to address these questions in subsequent data collection efforts. We also do not observe granular data on Fitbit app and device usage. However, this lack of usage data could attenuate our results (i.e., lack of use could drive our treatment effects closer to zero) as we are currently assuming all adopters to be using leaderboards in all periods after adoption. This makes our results more conservative.

Another potential limitation is that the student population in the sample may not be representative of the general population (e.g., they may be healthier, more physically active, or have more free time). Although broad generalization of results to the average population may be somewhat uncertain, it is useful to note that the largest leaderboard benefits accrue to the least active participants in our sample. These individuals are more comparable to the average population. More so, a younger population may be more amenable to gamification approaches and be more likely to be motivated by such interventions. Finally, we observe in our data the impact on physical activity and do not observe other downstream health outcomes (e.g., weight loss). However, given the documented relationship between physical exercise and other health outcomes and the magnitude of our effects, it is likely that individuals are, on average, healthier after leaderboard adoption.

These limitations notwithstanding, our results have significant implications for research and practice. Mitchell et al. (2013) argue for the potential of wearable technologies and mobile health more generally to be "leveraged to more promptly assess and reward behaviors on a population scale, further reducing the need for prohibitively costly incentives" (p. 666); however, there has been limited research validating this conjecture. We help substantiate this notion by demonstrating the significant potential for gamification-based interventions to meaningfully impact physical activity levels. Moreover, we find that the effects of leaderboards persist over time.

Our results also have implications for the general literature on motivating changes in health behavior. Specifically, gamification interventions coupled with health wearables provide notable advantages over other approaches studied in the literature. First, and unlike most other behavioral interventions, the widespread adoption of health wearables allows for much larger scale interventions. Second, because they are based in digital platforms, the design of these interventions can be tailored at the individual or group level to maximize benefit to the population. This customization is particularly valuable given the heterogeneity in benefit between subpopulations in our study and recognition by scholars of the limits of "one size fits all" approaches typically taken by most of the extant literature (Rogers et al. 2014, Rogers and Feller 2016).

Our research also highlights potential areas for future research. The variation in leaderboard effects points to the complex interplay between competition and social influence that underlies leaderboard effects and impacts motivation. Given that gamification approaches are highly varied, even for the same class of interventions (leaderboards can vary in terms of who can participate, who on the leaderboard is made salient to users, etc.), more work needs to be done to rigorously evaluate the potential benefits of these varied gamification approaches, the key mechanisms which drive their effects, and how they may have differential impacts across health contexts and individuals. Finally, there is a need to explore whether digital gamification interventions can be used in conjunction with classic interventions (decision aids, economic incentives, wellness programs, etc.) to unlock further health benefits.

Our results also have important implications for firms and policy makers. From the perspective of firms that sell health wearables and their related platforms, a key element of their value proposition is that their products positively impact health and well-being. Our results help substantiate the notion that physical activity can be positively impacted for large swaths of their adopters but also that some of their adopters (perhaps their earliest and most enthusiastic ones) can be harmed by some of the interventions they offer. Our results also suggests that these interventions need to be designed with careful consideration for how mechanisms intended to be motivational may not be so for all users. More so, this highlights the value of customizing gamification interventions to individuals and the value of nudging individuals toward a set of features that are likely to benefit them most. Thus, these firms may need to better incorporate insights from behavioral research in the design of gamification interventions available alongside health wearables.

Furthermore, policy makers, employers, and insurers are experimenting with health wearables and gamification as they have significant incentives to encourage more active lifestyles, which may improve health, improve worker performance, and lower healthcare costs. Scholars similarly suggest that gamification coupled with health wearables can be a cost-effective way to encourage increases in physical activity and to do so at scale (Mitchell et al. 2013). However, a salient concern with these approaches is that less healthy individuals, who are often of primary interest to policy makers and employers, can be demotivated when competing with their more active counterparts. Our results highlight that this concern may be unfounded or at least less salient, as sedentary individuals may benefit substantially from these approaches. Although the harm to highly active individuals is not ideal, some of these harms can be alleviated by tailoring leaderboards for these groups, and in net, these approaches are still likely to be valuable to employers and policy makers. A lingering challenge with reaping this value may be encouraging less healthy individuals to take up these interventions (e.g., join leaderboards), and employers and policy makers may need to invest in incentives to increase uptake for this subset of individuals to maximize potential value.

## Endnotes

[1] The World Health Organization (WHO) Constitution defines health as "a state of complete physical, mental and social well-being and not merely the absence of disease or infirmity."

[2] See www.who.int/news-room/fact-sheets/detail/physical-activity.

[3] See www.cdc.gov/chronicdisease/resources/publications/factsheets/physical-activity.htm.

[4] See www.lexico.com/en/definition/activity_tracker.

[5] See www.wired.com/story/science-says-fitness-trackers-dont-work-wear-one-anyway and www.fastcompany.com/3031324/why-your-company-should-think-twice-about-gamification.

[6] Descriptive results substantiate this conjecture and suggest that highly active individuals are much more likely to dominate smaller leaderboards compared with larger ones, that is, the smaller leaderboards do not seem to challenge these highly active individuals as they easily dominate them even with their reduced activity levels.

[7] See www.merriam-webster.com/dictionary/leaderboard.

[8] Although we chose a particular form factor and a particular vendor, which is arguably the market leader at the time of the study, the similarity of leaderboards across platforms means our results may be relevant to other platforms as well.

[9] Fitbit previously offered 30-day fixed-time leaderboards that, to our knowledge, have been discontinued.

[10] Mechanisms of individual accountability and changes to exercise reference points are distinct from competition mechanisms. For instance, I may have a friend or family member on my leaderboard who is not credibly competing with me (e.g., because their performance exceeds my own by a huge margin) but this individual can still reach out and hold me accountable for my exercise goals or impact my perception of what is achievable for me.

[11] Approximately 600 students were recruited but we are left with 516 students after excluding early dropouts and always adopters of leaderboards.

[12] Due to the coarseness of the survey data relative to the Fitbit data and some nonresponse in the sample, we use these controls for robustness checks.

[13] The adoption of Fitbit is not a central concern in our study as every user in our sample is a Fitbit user.

[14] Another method to increase the plausibility of the common trends assumptions is to use propensity score to achieve covariate balance across the treated and untreated individuals. We perform this propensity score–based adjustment as a robustness check in a later section.

[15] In Online Appendix C, we discuss GBM, IPTW, and IPTW use with DID. Furthermore, Table C.3 shows pre- and post-IPTW covariate balance, and Table C.4 and Figure C.1 explain the influence of covariates on leaderboard adoption.

[16] Please see Abadie and Cattaneo (2018) for a recent discussion on such examinations of the common trend assumption and Abadie and Dermisi (2008) for an example in a two-period setting.

[17] Therefore, $L_{i(t+0)}, \ldots, L_{i(t+9)}$ denote the weekly breakdown of the actual adoption, whereas $L_{i(t-9)}, \ldots, L_{i(t-1)}$ denote the placebo treatments in the weeks prior to actual adoption. Furthermore, $L_{i(t+9)}$ denotes period 9 and beyond, whereas $L_{i(t-10)}$ denotes period −10 and before.

[18] Please see apnews.com/article/2700956044de4517a471a47c3243078b.

[19] Because the average number of participants on leaderboards was relatively small (2.3), a binary indicator for being first was sufficient to capture the impact of rank. Results are consistent when we use a continuous measure of the prior week's rank.

[20] We focused on active users because the Fitbit leaderboard hides inactive users (i.e., with zero steps) from the view of the focal user and does not use them in the ranking presented to individuals.

[21] The labels "sedentary" and "highly active" are specific to our population and may not reflect average steps for sedentary or highly active individuals in the general population.

[22] Please see Online Appendix Table H.11 for similar analysis using full data and interaction terms.

## References

Abadie A, Cattaneo MD (2018) Econometric methods for program evaluation. *Annu. Rev. Econom.* 10:465–503.

Abadie A, Dermisi S (2008) Is terrorism eroding agglomeration economies in central business districts? *J. Urban Econom.* 64(2):451–463.

Bai Y, Hibbing P, Mantis C, Welk GJ (2018) Comparative evaluation of heart rate-based monitors: Apple Watch vs Fitbit Charge HR. *J. Sports Sci.* 36(15):1734–1741.

Charness G, Gneezy U (2009) Incentives to exercise. *Econometrica* 77(3):909–931.

Deci EL, Betley G, Kahle J, Abrams L, Porac J (1981) When trying to win: Competition and intrinsic motivation. *Personality Soc. Psych. Bull.* 7(1):79–83.

Deterding S, Dixon D, Khaled R, Nacke L (2011) From game design elements to gamefulness. *Proc. 15th Internat. Academic MindTrek Conf.: Envisioning Future Media Environ.* (ACM, New York), 9–15.

Festinger L (1954) A theory of social comparison processes. *Human Relations* 7(2):117–140.

Finkelstein EA, Haaland BA, Bilger M, Sahasranaman A, Sloan RA, Nang EEK, Evenson KR (2016) Effectiveness of activity trackers with and without incentives to increase physical activity. *Lancet Diabetes Endocrinology* 4(12):983–995.

Garcia SM, Tor A (2009) The N-effect: More competitors, less competition. *Psych. Sci.* 20(7):871–877.

Garcia SM, Tor A, Schiff TM (2013) The psychology of competition: A social comparison perspective. *Perspective Psych. Sci.* 8(6):634–650.

Hamari J, Koivisto J (2013). Social motivations to use gamification. *Proc. 21st Eur. Conf. on Inform. Systems*, 5–8.

Hamari J, Huotari K, Tolvanen J (2014a) Gamification and economics. Walz SP, Deterding S, eds. *The Gameful World* (MIT Press, Cambridge, MA), 139–162.

Hamari J, Koivisto J, Sarsa H (2014b) Does gamification work? *Proc. 47th Hawaii Internat. Conf. on System Sci.* (IEEE, New York), 3025–3034.

Handel B, Kolstad J (2017) Wearable technologies and health behaviors: New data and new methods to understand population health. *Amer. Econom. Rev.* 107(5):481–485.

Ho Y-JI, Liu S, Wang L (2022) Fun shopping: A randomized field experiment on gamification. *Inform. Systems Res.*, ePub ahead of print August 1, https://doi.org/10.1287/isre.2022.1147.

Jakicic JM, Davis KK, Rogers RJ, King WC, Marcus MD, Helsel D, Rickman AD, Wahed AS, Belle SH (2016) Effect of wearable technology combined with a lifestyle intervention on long-term weight loss. *JAMA* 316(11):1161–1171.

James TL, Wallace L, Deane JK (2019) Using organismic integration theory to explore the associations between users' exercise motivations and fitness technology feature set use. *Management Inform. Systems Quart.* 43(1):287–312.

Johnson D, Deterding S, Kuhn K-A, Staneva A, Stoyanov S, Hides L (2016) Gamification for health and wellbeing: A systematic review of the literature. *Internet Interventions* 6:89–106.

Lewis ZH, Lyons EJ, Jarvis JM, Baillargeon J (2015) Using an electronic activity monitor system as an intervention modality: A systematic review. *BMC Public Health* 15(1):1–15.

Liu C-W, Gao G, Agarwal R (2019a) *Reciprocity or Self-Interest? Leveraging Digital Social Connections for Healthy Behavior* (Social Science Research Network).

Liu C-W, Gao G, Agarwal R (2019b) Unraveling the "social" in social norms: The conditioning effect of user connectivity. *Inform. Systems Res.* 30(4):1272–1295.

Liu D, Santhanam R, Webster J (2017) Toward meaningful engagement: A framework for design and research of gamified information systems. *Management Inform. Systems Quart.* 41(4):1011–1034.

Lupton D (2016) *The Quantified Self* (John Wiley & Sons, Hoboken, NJ).

Mitchell MS, Goodman JM, Alter DA, John LK, Oh PI, Pakosh MT, Faulkner GE (2013) Financial incentives for exercise adherence in adults: Systematic review and meta-analysis. *Amer. J. Preventive Medicine* 45(5):658–667.

Pamuru V, Khern-am nuai W, Kannan K (2021) The impact of an augmented-reality game on local businesses: A study of Pokémon Go on restaurants. *Inform. Systems Res.* 32(3):950–966.

Patel MS, Asch DA, Volpp KG (2015) Wearable devices as facilitators, not drivers, of health behavior change. *JAMA* 313(5):459–460.

Patel MS, Volpp KG, Rosin R, Bellamy SL, Small DS, Fletcher MA, Osman-Koss R (2016) A randomized trial of social comparison feedback and financial incentives to increase physical activity. *Amer. J. Health Promotion* 30(6):416–424.

Penedo FJ, Dahn JR (2005) Exercise and well-being: A review of mental and physical health benefits associated with physical activity. *Curr. Opin. Psychiatry* 18(2):189–193.

Piwek L, Ellis DA, Andrews S, Joinson A (2016) The rise of consumer health wearables: Promises and barriers. *PLoS Medicine* 13(2):e1001953.

Rogers T, Feller A (2016) Discouraged by peer excellence: Exposure to exemplary peer performance causes quitting. *Psych. Sci.* 27(3):365–374.

Rogers T, Milkman KL, Volpp KG (2014) Commitment devices: Using initiatives to change behavior. *JAMA* 311(20):2065–2066.

Santhanam R, Liu D, Shen W-CM (2016) Gamification of technology-mediated training: Not all competitions are the same. *Inform. Systems Res.* 27(2):453–465.

Schultz PW, Nolan JM, Cialdini RB, Goldstein NJ, Griskevicius V (2007) The constructive, destructive, and reconstructive power of social norms. *Psych. Sci.* 18(5):429–434.

Shameli A, Althoff T, Saberi A, Leskovec J (2017) How gamification affects physical activity: Large-scale analysis of walking challenges in a mobile application. *Proc. 26th Internat. Conf. on World Wide Web Companion*, 455–463.

Steinhage AL, Cable D, Wardley DP (2015) Winning through cheating or creativity. *Academy of Management Proc.* vol. 2015.

Sullivan AN, Lachman ME (2017) Behavior change with fitness technology in sedentary adults. *Frontiers Public Health* 4:289.

Sun T, Gao G, Jin GZ (2019) Mobile messaging for offline group formation in prosocial activities: A large field experiment. *Management Sci.* 65(6):2717–2736.

Swan M (2013) The quantified self. *Big Data* 1(2):85–99.

Treiblmaier H, Putz L-M, Lowry PB (2018) Research commentary: Setting a definition, context, and theory-based research agenda for the gamification of non-gaming applications. *AIS Trans. Human-Comput. Interactions* 10(3):129–163.

Uetake K, Yang N (2019) Inspiration from the "biggest loser": Social interactions in a weight loss program. *Marketing Sci.* 39(3):487–499.

Wahl Y, Düking P, Droszez A, Wahl P, Mester J (2017) Criterion-validity of commercially available physical activity tracker to estimate step count, covered distance and energy expenditure during sports conditions. *Frontiers Physiology* 8:725.

Warburton DE, Nicol CW, Bredin SS (2006) Health benefits of physical activity: The evidence. *Canadian Med. Assoc. J.* 174(6):801–809.

Wu Y, Kankanhalli A, Huang KW (2015) Gamification in fitness apps: How do leaderboards influence exercise? *ICIS Proc.*

Xu Y (2017) Generalized synthetic control method: Causal inference with interactive fixed effects models. *Political Anal.* 25(1):57–76.